\documentclass[prb,showpacs,twocolumn]{revtex4}
\usepackage{amsmath}
\usepackage{graphicx}

\begin{document}

\title{A proposal to detect vortices above the superconducting transition temperature}

\author{Zoran Ristivojevic and M. R. Norman}
\affiliation{Materials Science Division, Argonne National Laboratory, Argonne, IL 60439}

\begin{abstract}
We propose a simple experiment to determine whether vortices persist above the superconducting transition temperature $T_c$
in the pseudogap phase of high temperature cuprate superconductors.  This involves using
a magnetic dot to stabilize a vortex in a thin cuprate film beneath the dot. We calculate the magnetic field profile as a function of distance from the dot if a vortex is
present, and discuss possible measurements that could be done to detect this.
Finally, we comment on the temperature range where a stable vortex should be observable.
\end{abstract}
\date{\today}
\maketitle

The superconducting order parameter is defined through its amplitude and phase. In conventional superconductors, phase fluctuations are relatively unimportant because of their large energy cost with respect to the Cooper pair binding energy. This ratio appears to be reversed for underdoped cuprates
as they are doped Mott insulators.\cite{Emery+95}
The reduced screening which results from this resembles the situation in thin superconducting films.\cite{Pearl64,Fetter+67,Halperin+79}

Vortices are topological objects defined by zeros of the order parameter that are created by winding of
the phase. Their existence in superconductors requires a well defined amplitude of the order parameter outside the vortex core, which is certainly satisfied below the transition temperature $T_c$. Above $T_c$ in conventional superconductors, the amplitude is non-zero in a very limited region of temperatures, whereas in cuprates this range appears to be much broader.  A variety of experiments \cite{Timusk+99} have suggested that this
amplitude might exist all the way to the pseudogap temperature $T^*$, a temperature much higher
than $T_c$ for underdoped cuprates. However, a non-vanishing amplitude is not a sufficient condition for vortices to be well defined.  For instance, in Kosterlitz-Thouless theory, the free vortex density above $T_{KT}$ rapidly increases with temperature.\cite{Minnhagen87}  Once the vortices begin to overlap at some temperature $T_L$, they become ill defined, and for temperatures above this it is expected that a gaussian picture for the fluctuations should be an adequate description.

This issue has received renewed significance with the observation of a large Nernst signal in the pseudogap phase of cuprates that extends
well above $T_c$.\cite{Xu+00}  Given that vortices are the origin of the large Nernst signal below $T_c$,
is their presence necessary to explain the large signal that persists above $T_c$?\cite{Ussishkin+04}
This is especially relevant given the strong
Nernst signal one estimates from a gaussian approximation for the superconducting
fluctuations.\cite{Ussishkin+02,Serbyn+09,Michaeli+09}
This question is further complicated by the observation that density wave reconstruction of the Fermi
surface in the pseudogap phase might give rise to a substantial part of the Nernst signal.\cite{Louis+09}

\begin{figure}[t]
\includegraphics[width=0.8\linewidth]{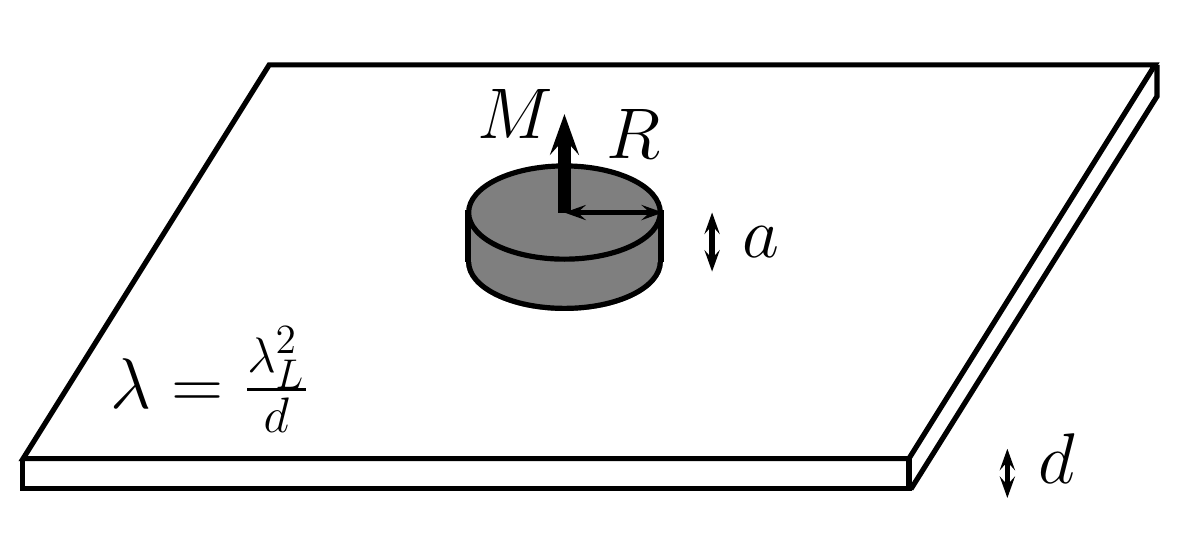}
\caption{Magnetic dot on a superconducting film with a
permanent magnetization $M$ perpendicular to the film.
$\lambda$ is the Pearl penetration depth of the film.} \label{Fig1}
\end{figure}

In this paper we propose a simple hybrid system \cite{Lyuksyutov+05,Velez+08} that could resolve the question about vortices in the pseudogap phase. This consists of a thin superconducting cuprate film and a small ferromagnetic dot placed on top of the film (Fig.~\ref{Fig1}).
The magnetization of the ferromagnetic dot generates a magnetic field that penetrates the superconductor. As a response to that field, supercurrents and (under certain conditions) vortices are induced in the film.\cite{Lyuksyutov+98}
The dot produces a potential well for the vortices. When this well is deep enough (i.e.~for large enough magnetization), and the  temperature is not too high, the
well can trap a vortex. The presence of the vortex can be
detected by measuring the magnetic field at the surface of
the film.

We now introduce the parameters of the hybrid system. We model the magnetic dot as a cylinder having a radius $R$, a height $a$ and a magnetization $M$ that is directed perpendicular
to the film (Fig.~\ref{Fig1}). The film has a London penetration length $\lambda_L$, a coherence
length $\xi$ and a thickness $d$.
When $d\ll \lambda_L$, the current density of the film is essentially uniform along the film thickness.\cite{Pearl64} Then the effective (Pearl) penetration length $\lambda=\lambda_L^2/d$ characterizes the magnetic response of the film. In the following we consider the situation where $\lambda\gg R>a$.
The magnetic field in the film is parallel to $M$ under the dot and becomes antiparallel outside. A sufficiently large $M$ can create a vortex underneath the dot,\cite{Lyuksyutov+98,Erdin+02} when the attractive vortex-magnet interaction overcomes the vortex creation energy. The interaction energy between a vortex (under the center of the dot) and the dot is\cite{Ristivojevic08}
\begin{align}\label{Umv}
U_{mv}=-Ma\phi_0\frac{R}{2\lambda},
\end{align}
where $\phi_0$ is the magnetic flux quantum. The interaction energy (\ref{Umv}) may be understood as a product of the magnetic moment of the dot $\mu=\pi M R^2 a$ and the average magnetic field produced by the vortex; the latter is equal to the flux of the vortex through the dot $\phi_0 R/(2\lambda)$ divided by the surface area $\pi R^2$ of the base of the dot.

The single vortex energy is
\begin{align}\label{Uv}
U_v=\frac{\phi_0^2}{16\pi^2\lambda}\ln\frac{L}{\xi}
\end{align}
under the condition that the typical film dimension $L$ is smaller than $\lambda$.
This is the limit of dirty superconducting films that have a large penetration length.\cite{Beasley+79} In the opposite case of large films, when $L\gg \lambda$, one would have $2.25\lambda$ instead of $L$ in Eq.~(\ref{Uv}).
Now the condition for the creation of a single vortex in the film is
\begin{align}\label{mu_c}
\mu\ge\mu_c=\frac{\phi_0 R}{8\pi}\ln\frac{L}{\xi},
\end{align}
which is obtained by requiring that the total energy $U_{mv}+U_v$ is negative.
We should emphasize that for magnetic moments much larger than $\mu_c$, there is a diversity of vortex-antivortex configurations that can be induced in the film.\cite{Ristivojevic08} In the following we are interested in the parameter range where only a single vortex exists under the dot, i.e.~we consider a magnetic dot with a magnetic moment near $\mu_c$.

\begin{figure}[t]
\includegraphics[width=0.5\linewidth]{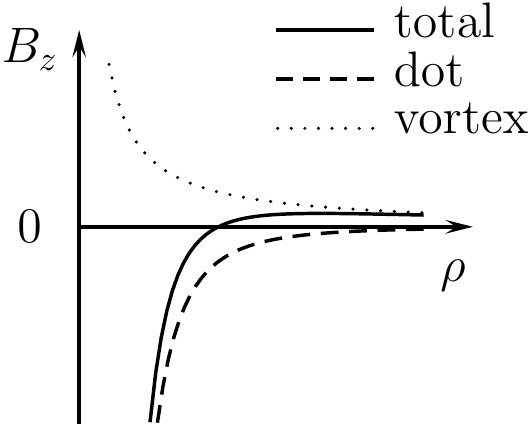}
\caption{The radial dependence of the perpendicular magnetic field along the surface of the film when a single vortex is induced underneath the magnetic dot. The total field (solid curve) is a sum of the contributions due to the dot (dashed curve) and the vortex (dotted curve).} \label{Fig2}
\end{figure}

The simplest way to detect the presence of a vortex would be to measure the total magnetic flux,
for instance by a scanning SQUID probe.\cite{Tafuri+04}
The additional contribution of the vortex for a typical sized SQUID loop would result in a
substantial fraction of $\phi_0$ ($\sim \phi_0/2$ for a loop radius of order $\lambda$).  The background flux associated with the dot would be easily
detected by going to a temperature significantly higher than $T_c$ (we assume that the magnetization
of the dot is $T$ independent for temperatures of order $T_c$).

More information can be obtained from the spatial profile of the magnetic field close to the film.\cite{Erdin+02}
The total magnetic field is a sum of contributions due to the field of the dot, the supercurrents and the vortex.
The contribution from the supercurrents is subleading for large $\lambda$. Therefore, the magnetic field perpendicular to the film at a distance $\rho$ ($a,R\ll\rho\ll\lambda$)
from the center of the dot along the film surface is (Fig.~\ref{Fig2})
\begin{align}
\label{B}
B_z(\rho)=-\frac{\mu}{\rho^3}+\frac{\phi_0}{4\pi\lambda}
\frac{1}{\rho}.
\end{align}
The first term is simply the dipolar magnetic field from the dot, while the second one is the field produced by the vortex. We see that the vortex acts to diminish the dot field. This can be exploited to detect
the presence of the vortex.  Given that the cuprates have a coherence length of order 20 $\AA$ and a London penetration depth of order 2000 $\AA$ [Ref.~\onlinecite{Hetel+07}], we would suggest a dot size with radius of order 200 $\AA$.  For a typical $\lambda$ of order a micron, then the additional
magnetic field due to the vortex near the edge of the dot would be around 100 G.
More interestingly, the total magnetic field at the film surface should change sign at a
distance $\rho_c=\sqrt{4\pi\lambda\mu/\phi_0}$,
or, taking the minimal magnetic moment that induces a single vortex (\ref{mu_c}),
\begin{align}
\rho_c=\sqrt{\frac{1}{2}\lambda R\ln\frac{L}{\xi}}
\end{align}
which has a value of order the London penetration depth.
Beyond this, the total field is opposite to that of the dot field, with a maximum at a
distance $\sqrt{3}\rho_c$.  The value of the field at this maximum scales like the inverse cube
of the London penetration depth.  For the parameters mentioned above, one obtains a field
value of about 2 G.
The change in the inhomogeneous field profile due to the vortex could potentially be detected from
a change in the nuclear magnetic resonance \cite{Vesna} or electron spin resonance lineshapes.
In this context, we note a recent experiment on cuprates that
detected an electron spin resonance signal above $T_c$ consistent with vortices.\cite{Talanov+10}
The presence of multiple vortices (and antivortices) for dots with $\mu > \mu_c$ will obviously
lead to a more complicated magnetic field profile.
One could also consider an array of dots for vortex trapping in cuprate films.\cite{Milosevic}

So far we have considered the zero temperature limit. At non-zero temperatures, thermal fluctuations can overcome the pinning potential of the dot. We model the the dot-vortex interaction by
\begin{align}
U_{mv}(\rho) = -\frac{\alpha}{\max(\rho,R)},
\end{align}
where $\alpha=\mu\phi_0/(2\pi \lambda)$, noting that $\lambda$ implicitly depends on $T$. Calculating the partition function of the vortex, we obtain the free energy \cite{footnote}
\begin{align}
F\approx U_v-T\ln\frac{L^2+R^2\exp\left(\frac{\mu\phi_0}{2\pi\lambda RT}\right)}{\xi^2}.
\end{align}
The previous equation defines a characteristic temperature
\begin{align}\label{Tf}
T_F=\frac{\mu\phi_0}{4\pi\lambda R\ln\frac{L}{R}}
\end{align}
with the following meaning: for $T\ll T_F$ thermal fluctuations do not essentially modify the zero temperature picture, while for $T\gg T_F$ thermal fluctuations overwhelm the pinning potential of the dot. Taking into account the critical magnetic moment of the dot which induces a vortex [Eq.~(\ref{mu_c})], we obtain $T_F=\phi_0^2/(32\pi^2\lambda)$ or $T_F=T_{KT}$, where $T_{KT}$ is the temperature for a Kosterlitz-Thouless (KT) transition in a thin film.\cite{Beasley+79,Doniach+79,Turkevich79,Halperin+79} Here we have made the reasonable assumption $L\gg R,\xi$. We have not taken into account the effects of bound vortex-antivortex pairs, but knowing that in the case of films without the dot they only weakly renormalize $\lambda$,\cite{Minnhagen87} we believe that their contribution in the present case can also be taken into account through a small renormalization of $\lambda$.

This now brings us to the question raised at the beginning, that is the existence of vortices above
$T_c$.  Terahertz conductivity experiments on underdoped cuprates have been successfully
modeled based on a dilute gas of vortices above $T_{KT}$.\cite{Corson+99,Orenstein+06}  These experiments
indicate that a finite bare phase stiffness temperature can be measured up to around 20K or so
above $T_c$.  This is the  limiting temperature, $T_L$, mentioned in the introduction, and it
corresponds to a vortex density times a core area
of order unity, with the vortex density smaller than this for $T < T_L$.\cite{Corson+99}
Since $T_L$ is considerably smaller than the expected $T^*$ based on the gap amplitude, the
vortices appear to be `fast' and `cheap' relative to classic superconductors.\cite{Lee+97,Ioffe+02}   What this means is that the core radius estimated from the above condition is about 400 $\AA$ as compared to the actual core radius of 20 $\AA$.\cite{Orenstein+06}  This implies a large `halo' around the vortex core
where the phase stiffness is reduced.  For our purposes, this dilute plasma limit \cite{Minnhagen87} is exactly the limit we want for seeing an isolated vortex under the dot (for a dot size of
order 200 $\AA$) that would not be perturbed too
significantly by the presence of free vortices above $T_{KT}$.  As the free vortex density depends
exponentially on temperature, diluting the free vortices to be even further apart to reduce their
perturbation still results in temperatures near $T_L$.
For $T > T_L$, the dilute
plasma limit is no longer valid, and the observation of `pseudo-vortices' by our proposed experiment
would not be possible.
The temperature range between $T_c$ and $T_L$ could be expanded by suppressing $T_c$,
for instance by stripe order.  This occurs for La$_{1.875}$Ba$_{0.125}$CuO$_4$, where indeed
strong KT signatures have been reported.\cite{Li+09}

This work was supported by the US DOE, Office of Science, under contract 
DE-AC02-06CH11357 and by the Center for Emergent Superconductivity, an Energy Frontier Research
Center funded by the US DOE, Office of Science, under Award No.~DE-AC0298CH1088.


\begin{thebibliography}{99}
\bibitem{footnote} Had we taken other forms for $U_{mv}$ that interpolate between the two limiting cases, $U_{mv}=-\alpha/R$ for $\rho \ll R$ and $U_{mv}=-\alpha/\rho$ for $\rho \gg R$, we would have obtained a similar expression. However the characteristic temperature $T_F$ given in Eq.~(\ref{Tf}) does not depend on these details to leading order.
\end{thebibliography}
\end{document}